\documentclass[acmlarge]{acmart}

\renewcommand\footnotetextcopyrightpermission[1]{} 
\usepackage{booktabs} 
\usepackage{balance} 
\usepackage{graphicx}
\usepackage{caption}
\usepackage{subcaption}
\usepackage{xcolor}
\usepackage{enumitem}

\begin{document}
\title{MFED: A System for Monitoring Family Eating Dynamics}
\thispagestyle{empty}

\author{Md Abu Sayeed Mondol}
\affiliation{
  \institution{University of Virginia}
}

\author{Brooke Bell}
\affiliation{
  \institution{University of Southern California}
}

\author{Meiyi Ma}
\affiliation{
  \institution{University of Virginia}
}

\author{Ridwan Alam}
\affiliation{
  \institution{University of Virginia}
}

\author{Ifat Emi}
\affiliation{
  \institution{University of Virginia}
}

\author{Sarah Masud Preum}
\affiliation{
  \institution{University of Virginia}
}

\author{Kayla de la Haye}
\affiliation{
  \institution{University of Southern California}
}

\author{Donna Spruijt-Metz}
\affiliation{
  \institution{University of Southern California}
}

\author{John C. Lach}
\affiliation{
  \institution{George Washington University}
}

\author{John A. Stankovic}
\affiliation{
  \institution{University of Virginia}
}

\begin{abstract}
 Obesity is a risk factor for many health issues, including heart disease, diabetes, osteoarthritis, and certain cancers. One of the primary behavioral causes, dietary intake, has proven particularly challenging to measure and track. Current behavioral science suggests that family eating dynamics (FED) have high potential to impact child and parent dietary intake, and ultimately the risk of obesity. Monitoring FED requires information about when and where eating events are occurring, the presence or absence of family members during eating events, and some person-level states such as stress, mood, and hunger. To date, there exists no system for real-time monitoring of FED. This paper presents MFED, the first of its kind of system for monitoring FED in the wild in real-time. Smart wearables and Bluetooth beacons are used to monitor and detect eating activities and the location of the users at home. A smartphone is used for the Ecological Momentary Assessment (EMA) of a number of behaviors, states, and situations. While the system itself is novel, we also present a novel and efficient algorithm for detecting eating events from wrist-worn accelerometer data. The algorithm improves eating gesture detection F1-score by 19\%  with less than 20\% computation compared to the state-of-the-art methods. To date, the MFED system has been deployed in 20 homes with a total of 74 participants, and responses from 4750 EMA surveys have been collected. This paper describes the system components, reports on the eating detection results from the deployments, proposes two techniques for improving ground truth collection after the system is deployed, and provides an overview of the  FED data, generated from the multi-component system, that can be used to model and more comprehensively understand insights into the monitoring of family eating dynamics.   
  
\end{abstract}



\maketitle

\section{Introduction}\label{sec:introduction}
Obesity increases one's risk for several health issues, including Type 2 diabetes, cardiovascular diseases, sleep apnea, osteoarthritis, kidney diseases, and certain cancers \cite{upadhyay2018obesity, niddk_risks}. While unhealthy dietary intake is one of the primary contributors to obesity risk, dietary intake has proven very difficult to measure and track. Research has shown that certain features of eating behaviors, contexts and events, such as when, where, with whom, and particular states of mind surrounding those eating behaviors, might be powerful determinants of food consumption and ultimately obesity\cite{lytle2010examining,boutelle2003associations, moens2009unfavourable}. With the rise of mobile technologies, possibilities have opened up to understand these eating behaviors more fully, in context and in real time. While the field has historically relied on self-report tools (such as 24-hour recalls, journals, and food frequency questionnaires \cite{12_short_dietry}) and more recently on image analysis of plated meals \cite{13_khanna2010overview}, all of these measurement modalities have substantial limitations in their accuracy to assess food intake. These measurement modalities can incur substantial participant burden, and/or research costs, for instance hours of coding, and cannot provide information in real time. Real time measurement will be important for future efforts that involve intervening just-in-time. Furthermore, these assessment tools often focus on the measurement of food intake independent of eating behaviors and the contexts in which eating occurs. Understanding these complex determinants of eating is an important, but understudied topic. The development of systems that can monitor and measure eating behaviors and their complex determinants in real time and in context is now possible with advances in wearable activity monitors. 

Family systems and family social networks (i.e., characteristics of the relationships and interactions among family members and emergent patterns of these interactions) are important dynamic milieus that impact eating. Empirical evidence shows that family members engage in similar food choices \cite{17_ayala2007association} and eating behaviors \cite{18_munsch2007restrained}, even across generations. Many dimensions of family systems can be barriers or promoters of healthy eating: family members are models for healthy or unhealthy habits \cite{15_pachucki2011social, 16_gorin2008weight} and family relationships can provide (or lack) information and support that influence eating behaviors and dietary intake \cite{17_ayala2007association}. Family relationships also influence, and are influenced by, personal states such as stress and mood; which may indirectly impact family members\textquotesingle \ eating. These strong links between family dynamics and eating highlight the potential to harness family influence to promote healthy eating and reduce disease risk, and mobile and wireless technologies offer new opportunities for measuring, tracking, and ultimately intervening upon these family eating dynamics (FED).

Real-time monitoring of FED requires detecting eating activities and related individual states and family features in the home, because this is a key environment in which family members interact, as well as prepare and consume food. This paper presents  MFED, a novel system for monitoring family eating dynamics in the home and in real-time. MFED uses smart wearables and Bluetooth beacons to monitor theoretically important features of family eating events and family dynamics while the users are at home. The main components of the systems are smartwatches, Bluetooth beacons, smartphones, a base station, and a cloud server. Smartwatches are used to detect when eating events are occurring in the home for each user. Bluetooth beacons are placed at different locations of the home to determine user location in the home. Smartphones are used to  measure individual states and situations (e.g., hunger, stress), via self-report on brief surveys, which are not captured by the sensor system. A base-station, placed at the home, collects and processes data from the watches and phones as well as manages the EMA (Ecological Momentary Assessment) surveys. The cloud collects data from all the homes and permits real-time monitoring of data being collected.
 
In contrast to traditional approaches that focus on individual\textquotesingle s dietary intake (e.g., what and how much is being eaten), our FED-based approach focuses on temporally dense and highly contextualized monitoring of family members\textquotesingle \ eating events while simultaneously capturing other theoretically-relevant states such as hunger, satiety, mood and stress, and family members\textquotesingle presence and interaction in the home. In the MFED system, eating events are monitored by detecting hand gestures for moving food or drink to mouth.  An ``eating event''  is a set of such gestures and represents phenomenon like consuming a meal, a snack, a drink or a combination of these consumption behaviors where eating gestures are clustered temporally. The MFED system is built for real-time detection of eating events at home so that EMA questions can be asked immediately after eating. It should be noted that our system monitors eating events, but it does not measure what people are eating. Other works have tried to measure what people are eating with varying success, and that knowledge has not resulted in improvements to the obesity problem. Instead, if we understand eating behaviors, new interventions might be possible to reduce obesity (showing this is outside the scope of this paper).

There are several challenges associated with developing a system like MFED. Some of the major challenges are discussed below.
\begin{itemize}
     \item \textit{Limited resources available in the smartwatch}: Detecting activities in real time using smartwatches is challenging as the resources (e.g., energy, computation, and memory) available in such devices are very limited. Continuous streaming of data to another device or to the base-station from a smartwatch consumes significant power from the watch. On the other hand, high performing methods like Convolutional Neural Networks (CNN)  require not only significant computation and memory, but also more energy to run the models.
     
    \item \textit{Detecting eating events in free-living context}: Detecting eating activities using wrist worn sensors, particularly in free-living context, is challenging \cite{zhang2017generalized}. The eating gestures and so the corresponding signals from the sensors differ widely for different foods, contexts, and utensils used as well as speed of moving hand to mouth. Confounding signals generated from an enormously wide range of non-eating activities make it difficult to detect eating gestures from a stream of sensor data. 
    
    \item \textit{Ground truth collection}: To evaluate the performance of the system in monitoring FED, and further improve it, ground truth is needed. However, in contrast to lab settings, it is challenging to collect ground truth in the wild once the system is deployed, particularly for systems like FED, partly because participants do not want additional intrusive devices in their homes (like cameras), and partly, if much later surveys are used they are error prone due to forgetfulness. 
    
    \item \textit{System installation}:  It is important to reduce the burden of installing a system and avoid intruding into the infrastructures (e.g., wiring, drilling etc.) of participants\textquotesingle \ homes. 
    
\end{itemize}

We developed innovative and effective approaches to address the challenges. Instead of streaming the sensor data continuously from the smartwatch to the server, we store the data in the smartwatch temporarily and detect potential eating events in the watch using an efficient approach. The data are uploaded to the base-station using Wi-Fi only when such an event is detected. The eating events on the smartwatch are detected conservatively to avoid false negatives. The data are further processed in the base-station with a more complex and effective method to finally detect the eating events. This approach reduces energy consumption from the smartwatch due to event-based data transfer instead of continuous transmission.

Once the potential eating event data is at the basestation we developed a two step solution. At first, we detect a set of potential eating gestures using a threshold based technique, and then we use a Convolutional Neural Network (CNN) to detect eating gestures from the set of potential gestures. The detected eating gestures are clustered together to detect an eating event. The algorithm is very efficient as most of the non-eating data are discarded using a simple threshold based method and only a small portion of the data is processed by CNN. Efficiency of our eating detection method allows MFED to use low-cost devices for base-stations. When an eating event is detected, the corresponding participant is asked to confirm the eating event immediately via a brief EMA survey, providing ground truth that is free from memory bias. In cases of correctly detected eating events, the participant is asked who they were eating with, among other questions, and response to this query provides ground truth for other participants who do not respond to the EMA or for whom the system fails to detect the eating event. This ``collaborative ground truth'' approach is novel for acquiring an improved level of ground truth in a free-living context.

MFED is very easy to install, and it is not intruding to the home infrastructures. Beacons are battery powered and they are attached in the walls using a removable mounting tape that does not damage paint or wallpaper. A beacon usually runs several years without battery replacement \cite{estimote}, and so the burden of maintenance is low even if the system is deployed for longer time. Apps for MFED can be easily installed in the smartwatches and the smartphones. Such a system is also very suitable for short and medium term deployment because no device needs to be installed permanently in the home. 

MFED provides real-time and contextualized information on eating events and allows us to monitor eating dynamics in families. The major contributions from this work are:

\begin{itemize}
\item We designed, developed and deployed MFED, a comprehensive system for monitoring family eating dynamics. To the best of our knowledge, this is the first system for monitoring FED.

\item We address the challenge of resource limitations of the smartwatches by an event based data upload method. 

\item We present a novel and efficient two-step algorithm for detecting eating events from wrist-worn accelerometers. The algorithm improves eating gesture detection f1-score by 19\%  with less than 20\% computation compared to the state-of-the-art methods. 

\item In addition to monitoring eating events, the system collects data on theoretically relevant features of family eating dynamics including: individual states (mood, stress, hunger/satiety) and characteristics of eating events (type of eating occasion, who they are eating with). 

\item We deployed the system for approximately two weeks in each of 20 real homes with a total of 74 participants. This paper provides preliminary summary results on eating activity from the deployments, and descriptions of the different features of FED that are being monitored.

\item We produce a dataset that consists of accelerometer, battery, and beacon readings from smartwatches as well as EMA responses from the users. The dataset will be useful in building better FED monitoring systems as well as new, dynamic and networked explanatory models of FED that can provide new insights into the unfolding of FED systems over time, and that can inform real-time, in the wild interventions in the future.

\item We present ``collaborative ground truth'', a novel approach for acquiring ground truth in free-living contexts and also use an hourly EMA as two techniques to improve ground truth collection in the wild.
        
\end{itemize}

\section{System Description}
\begin{figure}
\begin{center}
\resizebox{3.5in}{!}{\includegraphics{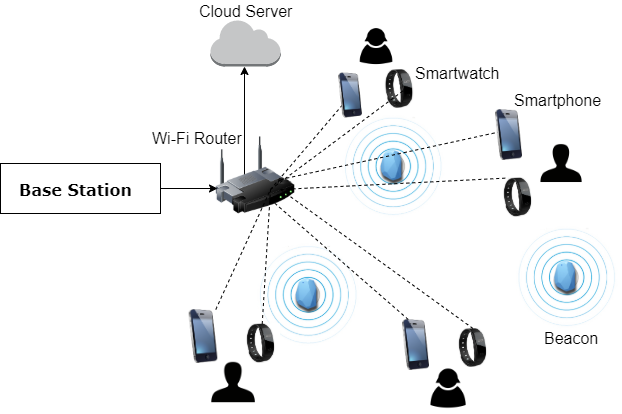}}
\end{center}
\caption{The MFED system mainly consists of smartphones, smartwatches, Bluetooth beacons, a base-station, and a cloud server. The smartphones, smartwaches, and the base-station are connected through a Wi-Fi router. The beacons broadcast Bluetooth packets that are scanned by the smartwatches.} 
\label{fig:overview}
\end{figure}

\begin{figure}
\begin{center}
\resizebox{6in}{!}{\includegraphics{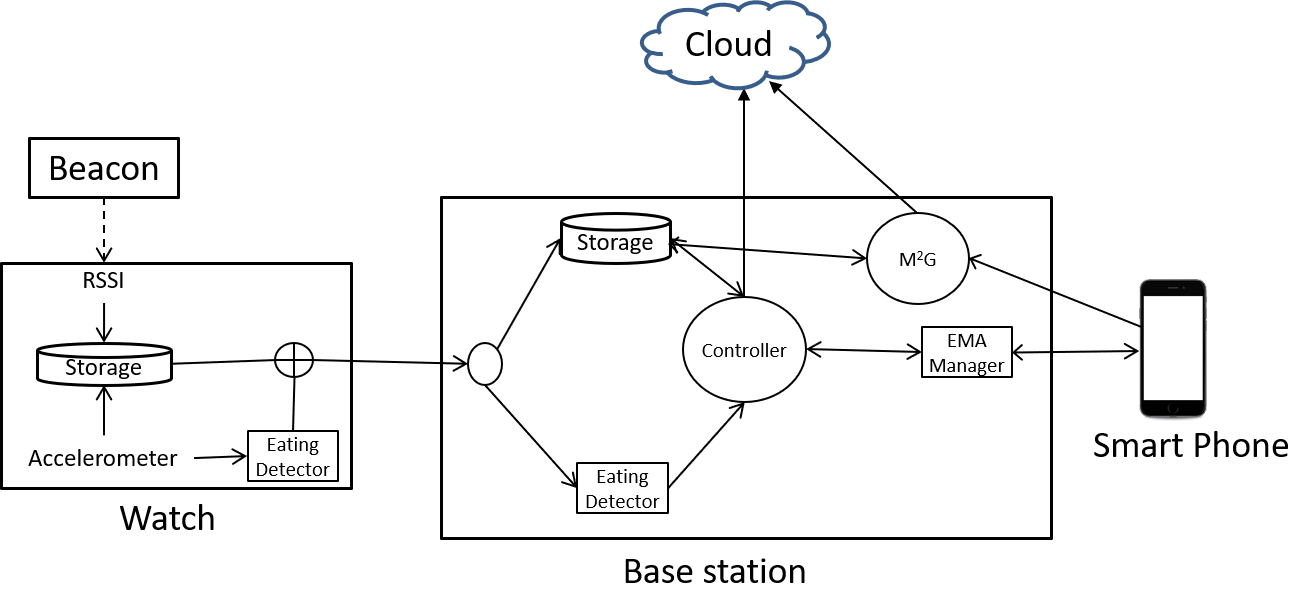}}
\end{center}
\caption{Dataflow in the MFED system} 
\label{fig:system_dataflow}
\end{figure}

The MFED system consists of smartwatches, smartphones, Bluetooth beacons, a base station, and a cloud server. Figure \ref{fig:overview} illustrates the connectivity among the devices in the system. The smartphones, the smartwatches, and the base station in the system are connected through a Wi-Fi router. Each of the family members uses a smartphone and wears a smartwatch. The beacons are placed at different locations in the home. Eating detection in MFED consists of two parts, one in the watch and the other in the base station. After an eating event is detected, an EMA is sent to the smartphone of the corresponding participant. To better understand the mood of the family members throughout the day (and not solely following detected eating events), the system deploys additional EMAs that include brief validated mood survey items several times over the day. The data flow in the system is depicted in Figure \ref{fig:system_dataflow}, and the following sections give the details of the different components of the system.    

\subsection{Sensors}
\subsubsection{\textbf{Beacon:}}
Bluetooth Low Energy (BLE) beacons are placed at different locations in the home. A beacon broadcasts packets that include the unique MAC address of the Bluetooth interface. The smartwatch scans for the Bluetooth packets and records the Received Signal Strength Indicator (RSSI) values of the signals that indicate the proximity of the watch to the beacons. There is no connection established between the beacons and the watches, rather it is a one-way communication where the beacons broadcast the packets independently, and any Bluetooth enabled device can receive the packet. The beacons broadcast continuously with several packets per second (a configurable parameter). We use Estimote \cite{estimote} beacons in our system. 

\subsubsection{\textbf{Smartwatch:}}
In our system, we use the Sony Smartwatch 3 \cite{sony_watch}, an Android-powered watch that has Wi-Fi capability. There were several reasons for choosing to use a wrist-worn sensor for eating detection. Though homes and utensils can be instrumented with sensors for detecting eating activity, this approach requires significant cost and effort for installation and maintenance. This approach is also limited in detecting the person who is eating, a major requirement for MFED. Video-based systems, often used for activity detection and person identification, are not suitable due to privacy issues. Also, such a system fails to detect when a person eats outside the camera view. Wrist devices such as smartwatches are ubiquitous and require very little effort for installation and maintenance. Because a wrist-worn device is a personal gadget, person identification using such a device is intuitive. Therefore, the use of smartwatches for eating detection in the MFED system appeared most advantageous. The Sony Smartwach 3 has several sensors including accelerometer and gyroscopes. The watch also has Blueotooth Low Energy (BLE) and Wi-Fi capability. 

\subsection{\textbf{Smartwatch App}}
We have developed an app for the smartwatch that collects data, detects potential eating events, and uploads the data to the base station in the background. The acceleromter data are collected and processed continuously when the watch is not being charged. The data are stored temporarily in the watch and uploaded to the base-station for further processing and actions whenever a sequence of potential eating gestures is detected. The details of detecting potential eating gestures are described in Section \ref{sec: eating_detection}. Since scanning RSSI signals from Bluetooth beacons consumes significant energy, the watch scans the beacon signals for 5 seconds with 2-minute interval to reduce energy consumption. The app also records the battery percentage of the watch once in the 2-minute interval. The battery and beacon RSSI data are uploaded opportunistically with the accelerometer data.

The watch app has Graphical User Interfaces (GUIs) which are used for configuration, testing and debugging. Once started, the app runs in the background without requiring any user  interaction. When the watch is restarted or turned on, the app starts automatically without any intervention. This is very critical for real-world deployments where users might forget to turn on the app when the watch is turned on. In fact, the user does not need to interact with the watch app at all. The GUIs are used by the research and deployment teams only.

\subsection{Base Station}
The base station has four major sub-systems: the Eating Event Detector, the EMA Manager, the MFED Controller, and the System Monitor. They are are described below. 

\subsubsection{\textbf{Eating Event Detector:}}
The base-station collects, processes and stores data from all the smartwatches used for MFED in the home. The eating event detector processed the accelerometer data and detects eating events in real-time. More detail about detecting eating events in the base-station is discussed in Section \ref{sec: eating_detection}. It should be noted that eating events are detected in the smartwatches just for uploading data to the base-station. The events detected in the base-station are ultimately used to send EMA surveys to the participants. 

\subsubsection{\textbf{Ecological Momentary Assessment (EMA) Manager:}}
There are two components of the EMA sub-system: The EMA Manager that runs in the base station, and the EMA App that runs in the smartphone.  The EMA Manager is responsible for sending EMAs to the smartphones, gathering the EMA responses from the smartphones, and storing the response data in the base station. The response data are eventually uploaded to the cloud by the MFED Controller.

\begin{figure}\centering
\includegraphics[width=\linewidth]{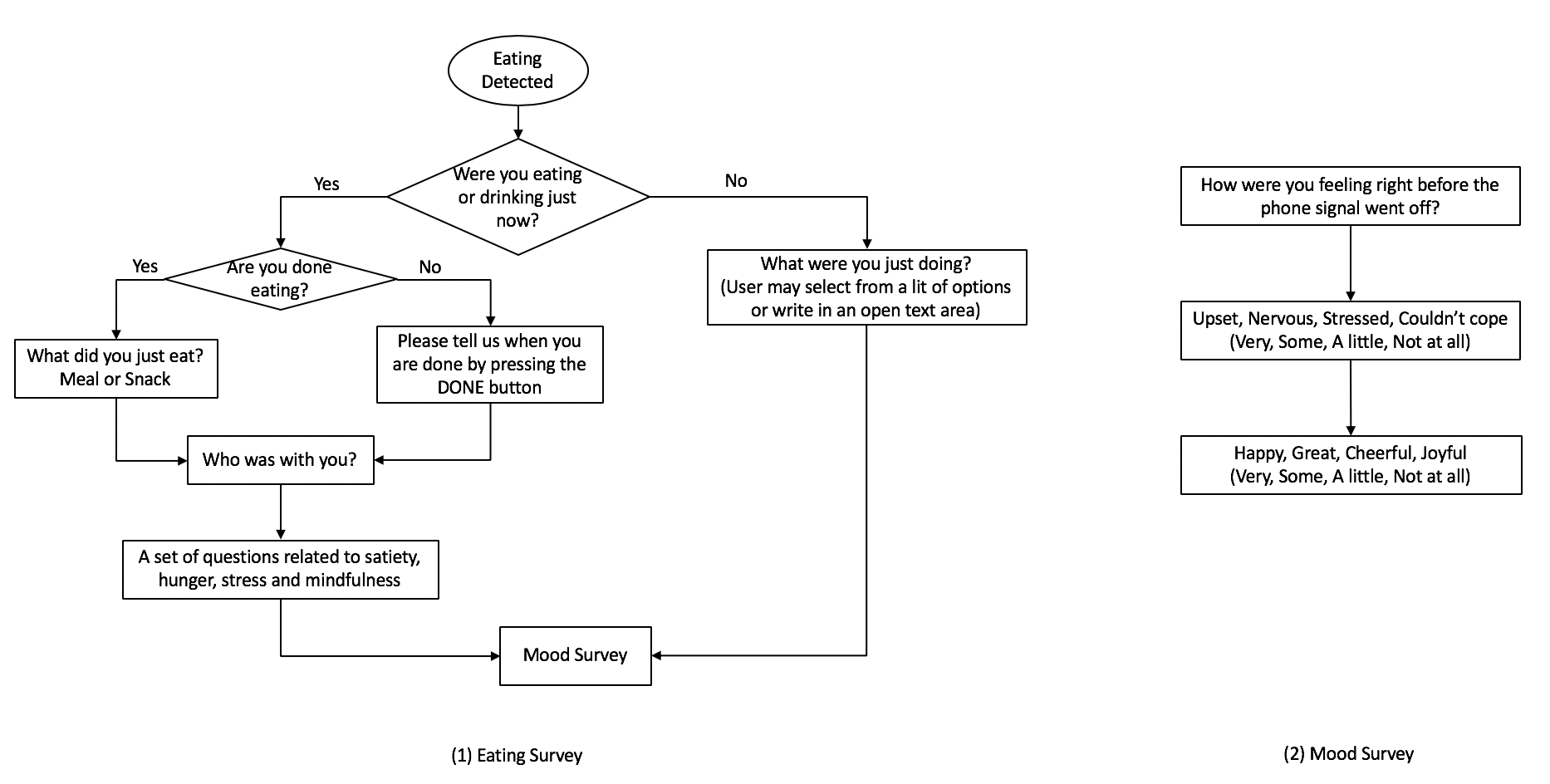}
\caption{EMA flowchart}
\label{fig:eating_ema_flowchart}
\end{figure}

There are two types of EMAs: eating EMA and mood EMA. An eating EMA is pushed out when an eating event is detected and a mood EMA that is pushed out according to a set time. To manage participant burden during the two week deployment, the system was set to send an EMA to participants no more than once per hour. This biases the system to the first eating event detected in the hour. So, if an eating event is detected at 12:15 and an eating EMA is sent out, the system will ignore an eating event detected at 12:45. If no eating event is detected in an hour, the mood EMA is sent out. The flow of the queries for the eating EMA is shown in Figure \ref{fig:eating_ema_flowchart} (1). At first, the participant is asked if he/she was eating. A negative response indicates that the eating event detected by the system is inaccurate. In such a case, the system asks the participant what he/she was doing and then a set of questions related to mood. If the detected eating event is accurate, the participant is asked to confirm whether he/she has finished eating. If eating is not finished, the participant is requested to press a DONE button available in the EMA app when finishing eating. When the eating event is finished, an EMA survey with validated measures of hunger, satiety, mindful eating, mood and stress is sent. The participant also provides information about who he/she was eating with, based on a multiple-response item. An eating EMA survey also includes questions related to mood.

As explained above, the mood EMAs are sent to the participants hourly unless an eating event is detected. The optimal frequency with which to assess stress and mood via ecological momentary assessment, with regard to both accuracy and compliance, is still an open question in the field. We chose 1-hour intervals to measure within-subject changes in mood and stress throughout the day. To account for the variation in daily routines and sleeping patterns of participants, we collect data about typical awake time from each participant before deployment. EMAs are sent only during this ``personalized participation window'' period. For example, if a participant\textquotesingle s window is from 6 am to 10 pm, EMAs are sent to that participant only during that time period. The participation windows are not same for all the participants of a home, rather they are set based on individual preference. The flowchart of mood EMA survey is shown in Figure \ref{fig:eating_ema_flowchart} (2). The survey includes validated items to assess key affective states: happy, joyful, upset, nervous, etc. More detail about the EMA survey is given in Section \ref{sec:individual_state}.

\subsubsection{\textbf{MFED Controller:}}
The MFED Controller is the integration module that runs continuously in the base station. The main purpose of the module is to start/stop the sub-systems, aggregate data from the sub-systems, and upload data to the cloud periodically. For each new deployment, it initializes all the parameters and starts the sub-systems. When a deployment is terminated, it stops the sub-systems and uploads remaining data from the base station to the cloud. The MFED Controller maintains communication among different modules. For example, when an eating event is detected, the Eating Event Detector sends the information to the MFED controller that triggers the EMA Manager to send EMA to the corresponding participant. 

\subsubsection{\textbf{System Monitor:}}
MFED is a system of systems, and it is necessary to monitor the status of the sub-systems in real time. For this purpose, we use $M^2G$ \cite{ma2017m}, a monitor for research-oriented residential systems. $M^2G$ monitors the processes of MFED running on the base-station,  the responses to the EMAs from the participants, and the battery status and network connection of the smartphones and smartwatches using the data uploaded to the base-station from these devices. In case of any discrepancy, it sends an alert via email to the deployment team responsible for addressing the issues. This type of real-time monitoring of the deployed system itself is important to minimize any loss of data due to various faults.


\subsection{Data Storage and Visualization}
The MFED controller uploads the data to a cloud server that stores all the data collected from the deployments. The continuous stream of data collected from an ongoing deployment is used to monitor the system and identify potential discrepancies or system failure. We have developed a web-based dashboard for visualization and interaction with the data from both ongoing deployments and past deployments. It displays the start time, end time, resident information, and devices for each resident of the deployment. For an ongoing deployment, the dashboard presents the critical information of the deployment including status of all the required processes (i.e., whether they are running properly or stopped at some point of time), the status of different data files that are supposed to be uploaded by the base station, and the network connectivity from the base station to the cloud. In addition, the dashboard also displays the latest battery status of the base station, the phones, and the watches used in that deployment. All this information is displayed with the corresponding timestamp. For example, if a process is not running, the dashboard shows the time when it stopped. The dashboard also provides an interface to visualize different data streams that correspond to different events like meals, EMA responses, mood, etc. For example, the list of EMAs sent to the participants as well the responses from them are accessible through the dashboard. An interface is also available to easily formulate a query to retrieve such data.

\section{Eating Detection} \label{sec: eating_detection}
\begin{figure}\centering
\includegraphics[width=0.8\linewidth]{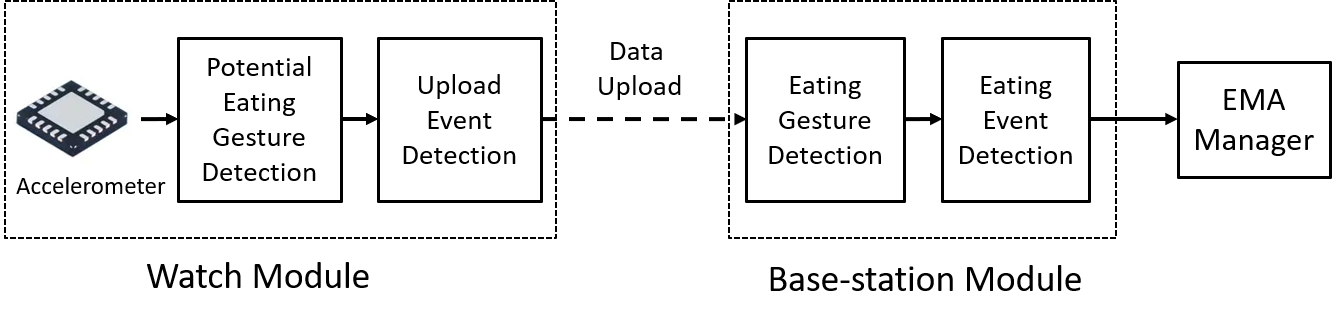}
\caption{The pipeline for eating event detection. This pipeline involves two modules: the watch module and the base-station module}
\label{fig:eating_pipeline}
\end{figure}

MFED detects eating in real-time at the base-station using the accelerometer data from the smartwatches. The pipeline for real-time eating detection consists of two modules: the watch module and the base-station module. The watch module collects data from accelerometer and uploads data to the base-station where eating events are detected. Instead of continuously streaming data to the base-station, the watch module stores data temporally in its memory, and then uploads the data to the base-station when a sequence of potential eating gestures are detected. The potential eating gestures are detected in the watch using a threshold based method, and it is used only to decide when the watch should upload data to the base-station so that the base-station can process the data further with more sophisticated algorithm, and can detect an eating event that triggers an EMA survey. The pipeline for eating event detection is depicted in Figure \ref{fig:eating_pipeline}. We describe these two modules in more depth below.

\subsection{Watch Module}
Though people may use one or both hands to eat food or drink, using watches on both wrists is not convenient, particularly in the wild. We place a watch only on the wrist of the dominant hand because this hand is generally used more for eating than the other hand \cite{zhang2017generalized, mirtchouk2017recognizing}. To develop the model for eating gesture detection, we collected data from 29 participants in lab settings where eating and other activities were not controlled.    

\subsubsection{\textbf{Potential Eating Gesture Detection: }}
\begin{figure}\centering
\includegraphics[width=0.7\linewidth]{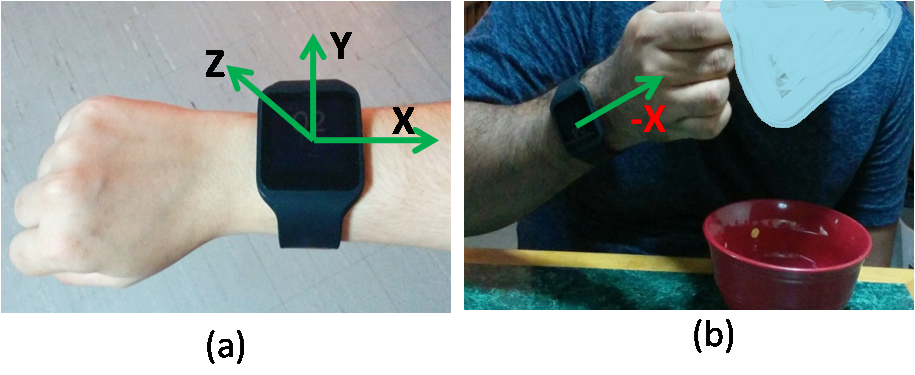}
\caption{(a) Axes of the sensors for an Android smart watches, (b) Hand orientation during an eating gesture}
\label{fig:watch_axes}
\end{figure}

Our potential eating gesture detection method is based on the fact that the wrist is usually inclined upward to some degree to move food to the mouth, and it goes down after that. We use the accelerometer from the wrist device to determine that the wrist is inclined upward. Accelerometers and other 3D inertial sensors are generally embedded in a wrist wearable device in such a way that one of the axes of the sensors is aligned with the arm length. For example, Figure \ref{fig:watch_axes}(a) shows the axes of an Android smartwatch worn on the right wrist, and the $X$ axis of the device is aligned with arm length. The alignment of the sensor axes with respect to the arm length might differ from device to device. Our method is also applicable for other arrangements. We use the arrangement of the Android smartwatch without loss of generality.

\begin{figure}\centering
\includegraphics[width=0.6\linewidth]{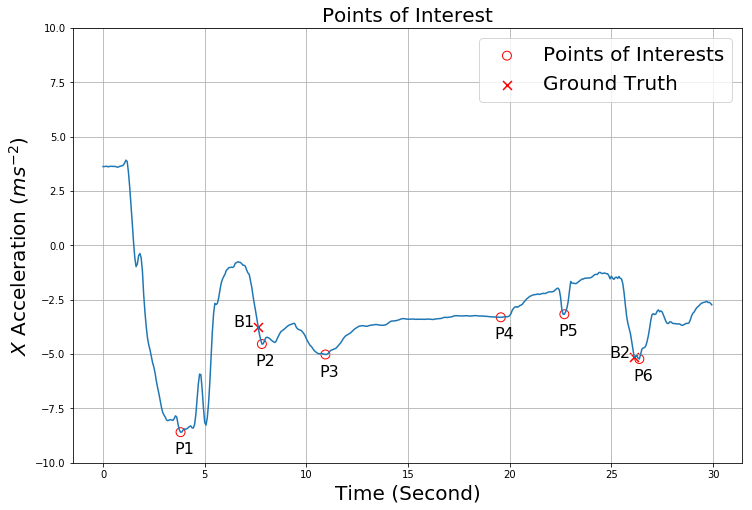}
\caption{Points of Interests}
\label{fig:poi_example}
\end{figure}

\begin{figure}\centering
\includegraphics[width=1.0\linewidth]{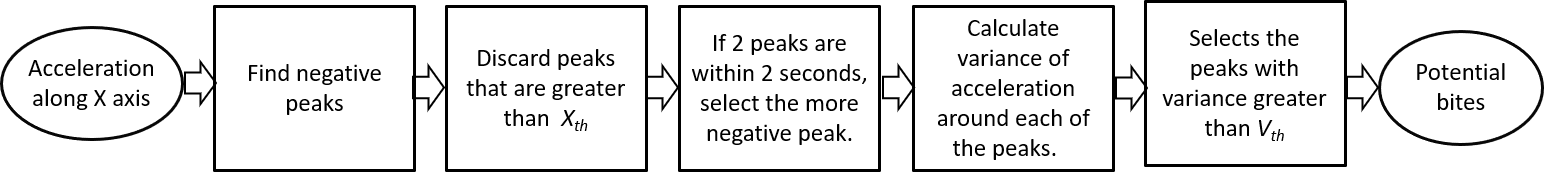}
\caption{Flowchart for potential eating gesture detection}
\label{fig:potential_eating gestures_flowchart}
\end{figure}

A value from an accelerometer is the total of the acceleration due to the gravity and the acceleration due to movement, called the gravitational acceleration and the linear acceleration, respectively. When the device doesn't move, the linear acceleration is zero, and so the accelerometer values represent the gravitational acceleration. The gravitational acceleration along the $X$ axis is negative while the wrist (i.e. the device) is inclined upward, and vice versa. Since the arm movements (and so the linear accelerations) are not usually intense during an eating gesture, the $X$ acceleration (the total of gravitational and linear acceleration) is  usually negative, particularly when the time series acceleration data is smoothed. The acceleration value generally decreases and then increases before and after an eating gesture, respectively. So, there is a negative peak in the acceleration data along the $X$ axis during an eating gesture. A point $P_i$ is defined as a negative peak if $AX_i < AX_{i-1}$ and $AX_i < AX_{i+1}$ where $AX_i$ denotes the acceleration value along the $X$ axis of the $i$-$th$ sample. This results in many peaks that are very close to each other. So, we use only the peak with the more negative value when two peaks are very close. From the data we collected, we found that most of the consecutive eating gestures are separated by more than 2 seconds, therefore we take the more negative peak when two consecutive peaks are within 2 seconds. Since the wrist is inclined upward to some degree during most of the eating eating gestures, the peaks with the $X$ acceleration value below a predefined threshold, called the $AX_{th}$, are selected. Finally, the peaks with very low or almost no movement are discarded, and the remaining peaks are considered as potential eating gestures. The degree of movement around a peak is measured by the variance of the acceleration around it. A window of 6 seconds is extracted around the potential eating gesture point, and the sum of the variances along each of the axes of the accelerometer is used for measuring the degree of movement. The peaks with variance greater than a threshold ($V_{th}$) remain as the potential eating gestures. The peaks are termed Point of Interest (PoI) in this paper. Figure \ref{fig:poi_example} shows an example of the $X$ acceleration with the PoIs for $X_{th}=-3$, and $V_{th}=1$. The number of potential eating gestures depends on the thresholds ($X_{th}$ and $V_{th}$), and more eating gestures are discarded from the potential eating gesture set when the threshold is lower. Importantly, the threshold values are determined empirically so that a significant amount of non-eating gestures is discarded while most of the true eating gestures are retained. Figure \ref{fig:potential_eating gestures_flowchart} shows the steps for potential eating gesture detection.

\subsubsection{\textbf{Upload Event Detection:}}
A sequence of the potential eating gestures indicates the possibility of an eating event. If four such potential eating gestures are detected in a 2-minute time window, data is uploaded from the watch to the base station for further processing and actions. The watch does not wait for 2-minutes, rather it uploads data as soon as four potential eating gestures are detected. The minimum interval between two data uploads from the watch to the base station is one minute. This reduces battery consumption from frequent data upload. However, it ensures that eating events are detected at the base station in real time which is required to trigger the eating EMAs. 

\subsection{Base-station Module}
\subsubsection{\textbf{Eating Gesture Detection:}}
Similar to the watch module, potential eating gestures are detected first from the accelerometer data. However, the threshold based method results many false positives. We use a Convolutional Neural Network (CNN) to detect eating gestures from the set of potential eating gestures. The length of the hand to mouth gestures for eating are not the same. Many of the windows we extracted for eating gestures includes non-eating gestures at the two ends. We can compare this with an image where the target object is somewhere in the image and does not necessarily cover the whole image. CNN has proven to be worked for image detection, and so we use CNN for eating gesture detection. CNNs have also shown to be effective in classifying time series data like those from wearable and smartphone sensors \cite{yang2015deep, zheng2014time}. 

The length of an eating gestures are usually less than 6 seconds \cite{thomaz2015practical, mirtchouk2017recognizing}, and we extract a window of this length around the PoIs.  An accelerometer provides data along X, Y and Z axes. So, the size of the window is $N \times 3$, where $N = 6 \times sampling\ rate\ of\ acceleration$. We reshape each window into $N \times 3 \times 1$ that resembles a 3-dimensional image. We use a filter of size $2 \times 2$ that results feature maps of width 2 and 1 in the first and the second convolution layer, respectively. A $2 \times 2$ filter convolutes two axes together in the first convolution layer, and thus captures the correlation between the axes, a feature widely used for accelerometer based activity recognition \cite{mondol2015harmony, mirtchouk2017recognizing}. The feature maps of width 1 in the second convolution layer limits further use of the filter, so we use two convolution layers in our solution. The smaller network also reduces the problem of over-fitting since the window of accelerometer data is much smaller in size than that of a typical image, and the number of instances in our dataset is much smaller than a typical image dataset.

\begin{figure}\centering
\includegraphics[width=0.5\linewidth]{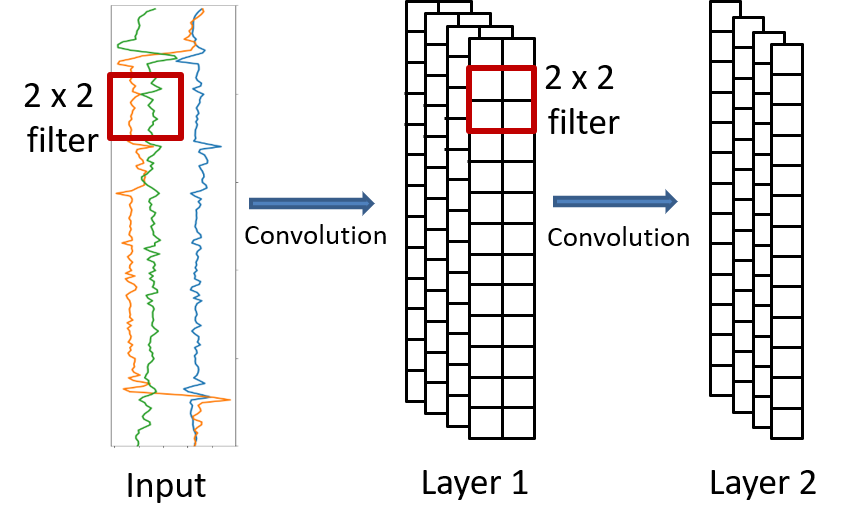}
\caption{Applying convolution to the acceleration along X, Y and Z axes. The width of the first and second convolution layers are 2 and 1, respectively.}
\label{fig:cnn}
\end{figure}

Figure \ref{fig:cnn} illustrates the network architecture for eating gesture detection. The input is processed by a Conv-Pool-Conv-Pool-Flatten-Dense-Dense network as depicted in Figure \ref{fig:cnn}. Each of the convolution layers is followed by a max-pooling layer. The output of the second pooling layer is flattened, and then two dense layers are applied. The final output is a single node that gives the probability of the input to be an eating gesture. ReLU (Rectified Linear Units) is used as activation functions for all the layers except the output layer where a sigmoid activation function is used. The output of the network is the probability of the input window being an eating gesture. Since it is a binary classification problem, we use a sigmoid activation function at the output layer. More details about the network with different parameter values and experimental results are given in \ref{sec:eating_results}.  


\begin{figure}\centering
\includegraphics[width=1.0\linewidth]{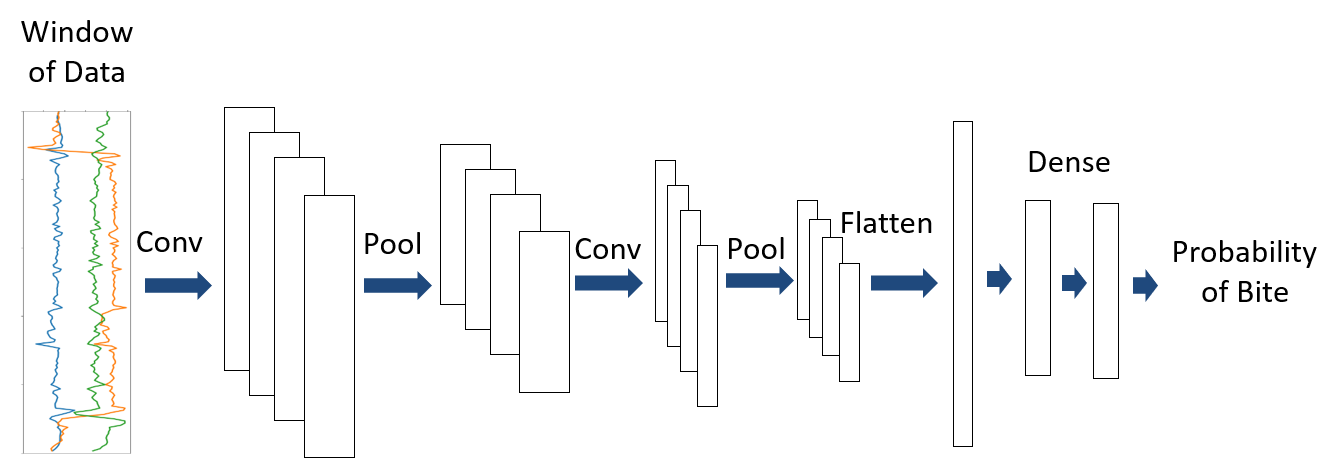}
\caption{The Convolutional Neural Network for eating gesture classification.}
\label{fig:cnn}
\end{figure}

\subsubsection{\textbf{Eating Event Detection:}}
\begin{figure}\centering
\includegraphics[width=0.7\linewidth]{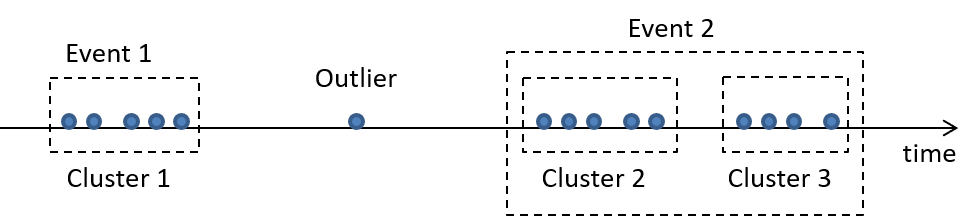}
\caption{Eating event formation from eating gestures.}
\label{fig:meal_cluster}
\end{figure}

An eating event is a sequence of eating gestures taken over some time and usually with irregular intervals. Like Mirtchouk et al. \cite{mirtchouk2017recognizing}, we construct an eating event by clustering eating gestures where two eating gestures within one minute-distance belong to the same cluster. However, if the interval between two detected eating gestures from an eating event is greater than one minute, more than one cluster might be generated for that single event. Therefore, clusters that are within four minutes distance (the waiting time before sending an EMA after an event is detected) are combined together. So, a single cluster of eating gestures or a cluster of the clusters define an eating event. Figure \ref{fig:meal_cluster} illustrates eating event formation from eating gestures where Event 1 is constructed from a single cluster and Event 2 is from 2 clusters. There might be some outlier eating gestures that are not part of any cluster or eating event. Such outliers are often generated by confounding gestures that are misclassified by the eating gesture detection method. It should be noted that an eating event can be detected even if some eating gestures are not detected by the system. On the other hand, some false eating gestures close to each other might form a cluster, resulting in a false eating event. Any cluster having at least 3 eating gestures is used for eating event detection. It reduces false positives while detecting events with fewer eating gestures.

\section{Individual State Detection} \label{sec:individual_state}
In addition to information regarding the detected eating events, MFED collected self-reported individual states including hunger, satiety, negative and positive affect, stress, and eating in the absence of hunger. There is no current solution available to detect these states automatically, and developing such a solution is beyond the scope of this work. We utilized Ecological Momentary Assessment (EMA)\cite{shiffman2008ecological}, a method of sampling users\textquotesingle \ behaviors and states in real time, to collect information from the user. The mood and eating EMAs included a set of queries assessing positive and negative affect and stress, while the eating EMAs additionally assessed hunger, satiety, and eating in the absence of hunger. The EMA survey questions are selected based on previous studies that have validated their usefulness in capturing the corresponding states. 

\begin{itemize}
    \item \textbf{Positive/Negative Affect and Stress}: An adapted\cite{laurent1999measure, cohen1983global} 8-item survey was used to assess users\textquotesingle \ momentary positive and negative affect/stress right before the phone survey was received. Users were asked to rate their level of being happy, great, cheerful, joyful (positive affect), upset, nervous, stressed, and couldn\textquotesingle t cope (negative affect/stress) on a 4-point Likert scale ranging from ``not at all'' (1) to ``very'' (4).\\
    
    \item \textbf{Hunger and Satiety:} Users were asked to rate how hungry they were right before they ate (hunger), and how full they were right after they ate (satiety), on a scale from 0 to 100 \cite{cardello2005development}.\\
    
    \item \textbf{Eating in the Absence of Hunger:} A 16-item survey \cite{cohen1983global} was used to assess the level at which users were eating in the absence of hunger. They were asked to consider reasons why they started eating, and reasons why they kept eating, such as ``food looked, tasted or smelled so good'' and ``my family or parents wanted me to eat'' and ``feeling sad or depressed''. Item responses ranged from ``not true'' (1) to ``very true'' (4).
\end{itemize}

\section{Collaborative and Hourly EMA-based Ground Truth}
In lab settings or initial in-home tests, ground truth of activities of the participants can be acquired from recorded video or real-time observation. However, once deployed the goal is to not add any extra burdens or equipment to obtain ground truth. Consequently, ground truth is usually unknown. To mitigate this issue, ground truth in the wild is generally collected from self-report significantly after the event. This is error prone. Another approach used in some studies \cite{thomaz2015practical, zhang2017generalized} is to use on-body cameras for ground truth in the wild. But this approach is often not acceptable to participants due to privacy concerns. Further, even if participants agree, they can easily forget to wear them. This raises a key issue that precise ground truth for realistic, long term deployments is not really possible.

For MFED, we significantly improve ground truth by using two new methods that require no new hardware beyond what the system needs for its original functionality. We refer to these solutions as collaborative and hourly based EMA ground truth. While these techniques can significantly improve obtaining ground truth and identifying false positives, the fact that our eating detection solution missed that some eating events occurred may still remain undetected. 

 In MFED, we ask a participant for confirmation when an eating event is detected for that participant. The confirmation not only leads to following EMA questionnaires for the participant, but also serves as ground truth for the corresponding eating events. This first-person approach of ground truth collection does not work if the system fails to detect the eating event or the participant does not respond to the EMA survey. Considering that people often eat at home with other family members, we can collect ground truth for the eating event of one person from another family member when they eat together and the later person response to his/her eating EMA survey. We denote this approach as ``Collaborative Ground Truth''. 

\begin{figure}\centering
\includegraphics[width=0.7\linewidth]{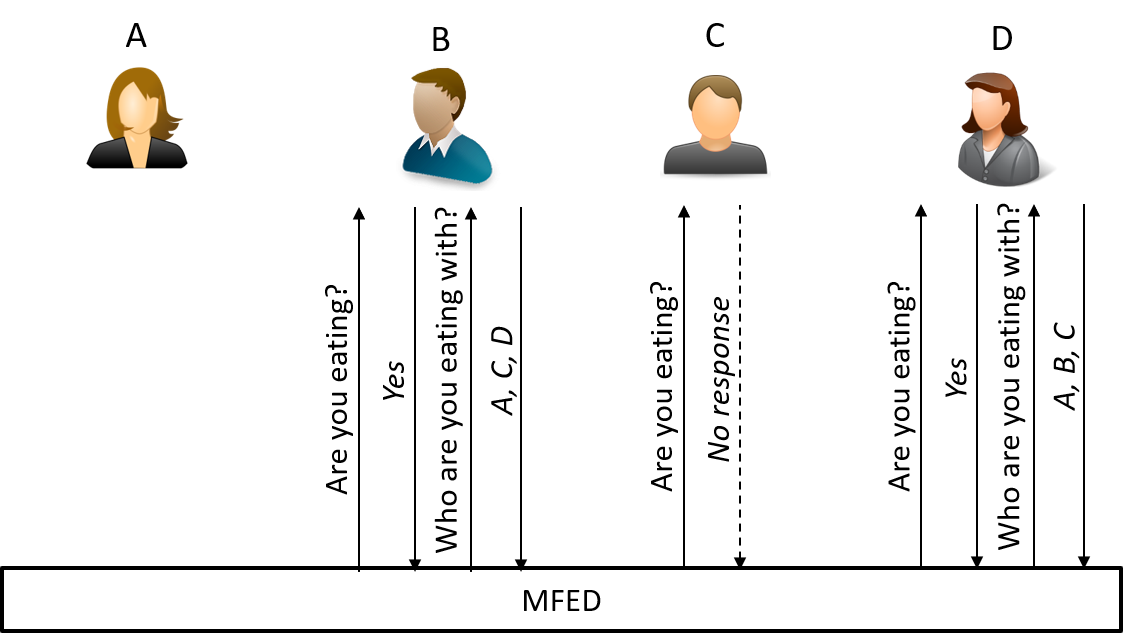}
\caption{A scenario when four family members (A, B, C and D) eat together. MFED detects eating events for B, C and D. C doesn't respond to the eating EMA survey. The responses from B or D can be used to collect ground truth for A and C.}
\label{fig:cgt}
\end{figure}

This works as follows: When a participant confirms a detected eating event, the participant is asked about who he/she was eating with. The response to this query provides ground truth for other participants. Figure \ref{fig:cgt} presents a scenario where four family members (A, B, C and D) eat together. MFED detects eating events for B, C and D, and sends an EMA survey to them. However, C doesn't respond to the eating EMA survey. However, B and D confirms that they are eating, and thus provides ground truth for their corresponding eating events. Following  their confirmation, B and D are asked who they are eating with. The responses from them can be used to collect ground truth for A and C. Such ground truth is also useful for localization as the two participants are co-located during the eating event.   

The second technique also utilizes EMA, but is based on hourly EMAs. Since MFED is interested in the mood of family members, it asks hourly questions about mood and for ground truth purposes we add a question, "did you eat in the last hour." So, if we failed to detect an eating event and the participant indicates that they did indeed eat we can detect a false positive for our eating detection solution. In this ground truth solution, the question is near in time to the event, making it less error prone, and it piggybacks with minimal burden on questions that are fundamental to the purpose of the study.

The effectiveness of these two techniques is shown in the experiments section.


\section{Experiments and Results}
\subsection{Eating Gesture Detection (Lab)} \label{sec:eating_results}
Prior to real deployments, we collected data from 29 participants for a total of 42 sessions in lab settings to develop a model for eating gesture detection. The average duration of the sessions was about 24 minutes. The participants ate freely while being video recorded for ground truth purposes. The videos were annotated using the ChronoViz \cite{chronoviz} tool.  Almost all the participants in our study are right-handed, and so we use data from the right hand only. The participants ate freely during the sessions and engaged with different non-eating activities including reading books, moving around, using phones, and using computers. To capture more activities, we collected data for non-eating activities from free living context involving 4 persons. The total duration of the free-living data is about 18 hours which includes different household activities except eating. We did not video record during these free living sessions since there was no eating activity present there.  

The moment when the food or drink reach to mouth is annotated, and an window around that moment is considered as an eating gesture. There are usually some time difference between a moment annotated and the PoI detected from the data for that eating gesture. In order to address this issue and reduce ambiguous instances, we label the potential eating gestures as positive, negative and ambiguous for training purpose. The potential eating gestures within 2 seconds of an annotation are labeled as positive (eating gestures). Those in the range of 2-4 seconds are labelled as ambiguous, and others as negative (non-eating gestures). The ambiguous potential eating gestures are not used for training, but all the potential eating gestures are used for testing the performance of the classification model. In order to determine the True Positives (TP), the False Positives (FP), and the False Negatives (FN), we use the method presented by Yujie et al. \cite{dong2012new} which is suitable for eating gesture detection, particularly when there are time differences between the annotated moments and the corresponding detected eating gesture moments. We take a  window of length 6 seconds around the potential eating gestures because most eating gesture gestures are captured well by a 6 second long window \cite{thomaz2015practical, mirtchouk2017recognizing}. We use 32 and 64 filters in the first and second convolution layer, respectively and 100 nodes in each of the dense layer. The experimental results used to select the filter numbers are explained later in this section.  We evaluated our method using a leave one person out approach where data from each person is tested by the model developed using data from other persons. 

\begin{figure}\centering
\includegraphics[width=0.5\linewidth]{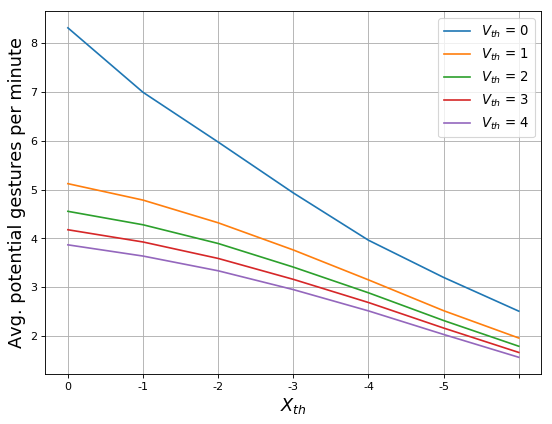}
\caption{ Average potential eating gestures per minute  for different values of $V_{th}$ and $X_{th}$}
\label{fig:avg_pbpm}
\end{figure}
\begin{figure}
\centering
\begin{subfigure}{.5\textwidth}
  \centering
  \includegraphics[width=.9\linewidth]{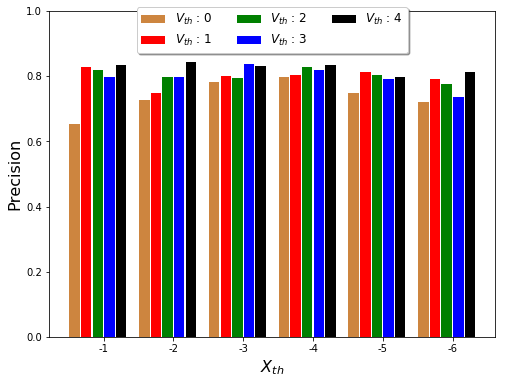}
  \caption{Precision}
  \label{fig:sub1}
\end{subfigure}%
\begin{subfigure}{.5\textwidth}
  \centering
  \includegraphics[width=.9\linewidth]{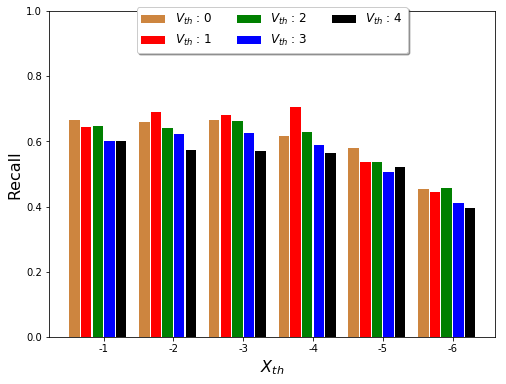}
  \caption{Recall}
  \label{fig:sub2}
\end{subfigure}
\caption{(a) Precision and (b) Recall of eating gesture detection for different values of $V_{th}$ and $X_{th}$}
\label{fig:pr_rc_eating gesture}
\end{figure}
\begin{figure}
\centering
\begin{subfigure}{.5\textwidth}
  \centering
  \includegraphics[width=.9\linewidth]{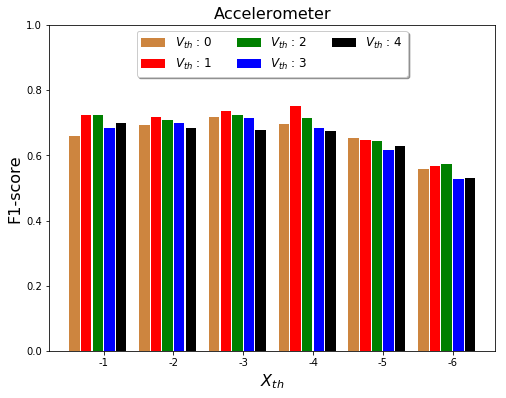}
  \caption{}
  \label{fig:sub1}
\end{subfigure}%
\begin{subfigure}{.5\textwidth}
  \centering
  \includegraphics[width=.9\linewidth]{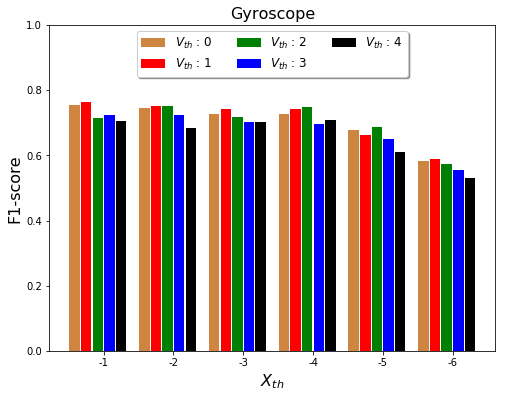}
  \caption{}
  \label{fig:sub2}
\end{subfigure}
\caption{F1-score of eating gesture detection (a) using an accelerometer (b) using both an acceleromter and a gyroscope.}
\label{fig:f1_eating gesture}
\end{figure}

Figures \ref{fig:pr_rc_eating gesture}(a), \ref{fig:pr_rc_eating gesture}(b) and \ref{fig:f1_eating gesture}(a) show the precision, recall and F1-scores of eating gesture detection for different threshold values for $X_{th}$ and $V_{th}$ that represent acceleration along the $X$ axis and the total variance  of acceleration, respectively.  The average numbers of potential eating gestures per minute are shown in Figure \ref{fig:avg_pbpm}. The number of potential eating gestures is much smaller for $V_{th}=1$ than $V_{th}=0$. However, the number decreases less significantly for further increments of $V_{th}$. The figure shows that the number of potential eating gestures decreases almost linearly with $X_{th}$. The precision values do not differ significantly for different values of $V_{th}$ and $X_{th}$. However, the recall value decreases significantly particularly for $X_{th}<-4$. This is because the more $X_{th}$ decreases, the more true eating gestures are discarded along with non-eating gestures. Figure \ref{fig:f1_eating gesture}(a) shows that F1-scores are less or similar when $X_{th}>-3$. However, it decreases significantly for $X_{th}<-4$. We found that the F1-score reduces significantly for some participants for $X_{th}=-4$. So, we selected 1 and -3 for $X_{th}$ and $V_{th}$, respectively. The F1-score of our method at these thresholds is witn

\begin{figure}\centering
\includegraphics[width=0.5\linewidth]{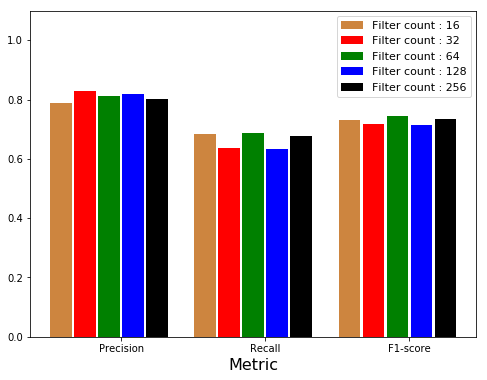}
\caption{Precision, recall and F1-score for different number of filters. Same number of filters are used in both layers.}
\label{fig:prec_recall_f1_kernel}
\end{figure}

The results for using different number of filters in the convolution layers are shown in Figure \ref{fig:prec_recall_f1_kernel} where we have used same number of filters in both the layers. There is no significant difference in the results for the different filter counts. To reduce the problem of over-fitting on the free-living data, we use 32 and 64 filters in the first and second convolution layers, following the approach of AlexNet\cite{krizhevsky2012imagenet} that uses fewer filters in the earlier layers than the following layers. In addition to the accelerometer, we also collected data from a gyroscope of the watch in the lab study. Figure \ref{fig:f1_eating gesture}(b) shows the F1-scores of eating gesture detection when a gyroscope is used in addition to the accelerometer. There is no significant difference between using an acceleromter only and using both an accelerometer and a gyroscope. However, a gyroscope consumes significant energy, usually more than an accelerometer  \cite{park2011gesture, katevas2016sensingkit}. So, we considered not to use the gyroscope for the in-home deployments.  

The computation required for the method used for potential eating gesture detection is negligible compared to the methods used for further classification. The sliding window based approach, used widely in state of the art solutions \cite{mirtchouk2017recognizing, thomaz2015practical}, segments the data usually with an overlap between consecutive segments. The number of segments depends on the sliding length. For example, Thomaz et al. \cite{thomaz2015practical} segments the data into 6 second long windows with 3 second overlap resulting 20 segments per minute. Mirtchouk et al. \cite{mirtchouk2017recognizing} segments the data into 5 second window with 100 millisecond step size resulting 600 segments per minute. Compared to the state of the art solutions, we use a very small number of segments for classification. For example, with $X{th}=-3$ and $V_{th}=1$, the average number of potential eating gestures (segments) per minute is 3.76, which is less than 1\% and 20\% compared to  the methods of Mirtchouk et al. and Thomaz et al., respectively. This is a significant reduction of computation requirement, and it makes our method more suitable for on-device and real-time processing, particularly for devices that have limited resources. Computation can be reduced further by decreasing $X_{th}$ or increasing $V_{th}$, but compromising performance to some extent (e.g. using $X_{th}=-5$). Figure \ref{fig:f1_eating gesture} shows that the thresholds we selected ($X_{th}=-3$ and $V_{th}=1$) gives similar or even better results than greater values of $X_{th}$ or smaller values of $V_{th}$ that result more potential eating gestures. Our method reduces computation requirement without any compromise of performance.  

\subsection{In-Home Deployments}
We recruited families that have at least one adult parent and one child between the ages of 11 and 18 years old living in Los Angeles, California. Children under the age of 11 were not eligible to participate in the study. We deployed the system in 20 homes with a total of 74 participants. Thirty-nine children (Average age = 15.3), fourteen adult males (Average age = 44.7), and twenty-one adult females (Average age = 45.0) were included in the sample. The majority of participants identified as Hispanic or Latino (64.9\%); 14.9\% identified as White or Caucasian, 10.8\% as mixed race, and 9.4\% as other. Table \ref{table:deployment_summary} shows the number of participants from each of the families. For some families, all members were not interested in participating in the study. However, mothers from all the families participated in the study. The number of participants from a family ranges from 2 to 5. 

We provided each of the participants with an Android smartwatch and an Android smartphone during the deployment period. The participants were instructed to primarily use the phone for study purpose. Because the scope of the study focused on in-home family eating dynamics, the participants were told to wear the watch only at home and during active daytime. Beacons were not placed in bedrooms or bathrooms to preserve the privacy of the participants. The system is deployed in each home for about 2 weeks.  We have two protocols approved by Institutional Review Board (IRB) to collect the lab data and to deploy the system at homes, respectively.

\begin{table}
\begin{tabular}{|c|c|c|c|c|c|c|}
\hline
\textbf{Family ID} & \textbf{Adult Female} & \textbf{Adult Male} & \textbf{Children}\\ \hline
1   & Y &   & YY             \\ \hline     
2   & Y & Y & Y              \\ \hline     
3   & Y & Y & Y              \\ \hline     
4   & Y & Y & YY             \\ \hline     
5   & Y & Y & Y              \\ \hline     
6   & Y & Y & YY             \\ \hline     
7   & Y & Y & YY             \\ \hline     
8   & Y & Y & YY             \\ \hline     
9   & Y &   & YY             \\ \hline     
10  & Y &   & YYYY            \\ \hline     
11  & Y & Y & YY              \\ \hline     
12  & Y &   & YY              \\ \hline     
13  & Y & Y & YYY            \\ \hline     
14  & Y &   & YY              \\ \hline     
15  & Y & Y & YY              \\ \hline     
16  & Y &   & Y               \\ \hline     
17  & Y & Y & YYY             \\ \hline     
18  & Y & Y & YY              \\ \hline     
19  & YY& Y & Y             \\ \hline     
20  & Y & Y & YY              \\ \hline     

\end{tabular}
\caption{Number of participants with different age groups in the families. The count of 'Y' in a cell indicates the number of participants in the corresponding age group from the corresponding family.}
\label{table:deployment_summary}
\end{table}

\subsubsection{\textbf{EMA Responses:\\}}
\begin{figure}
\centering
\begin{subfigure}[b]{.9\textwidth}
  \centering
  \includegraphics[width=.9\linewidth]{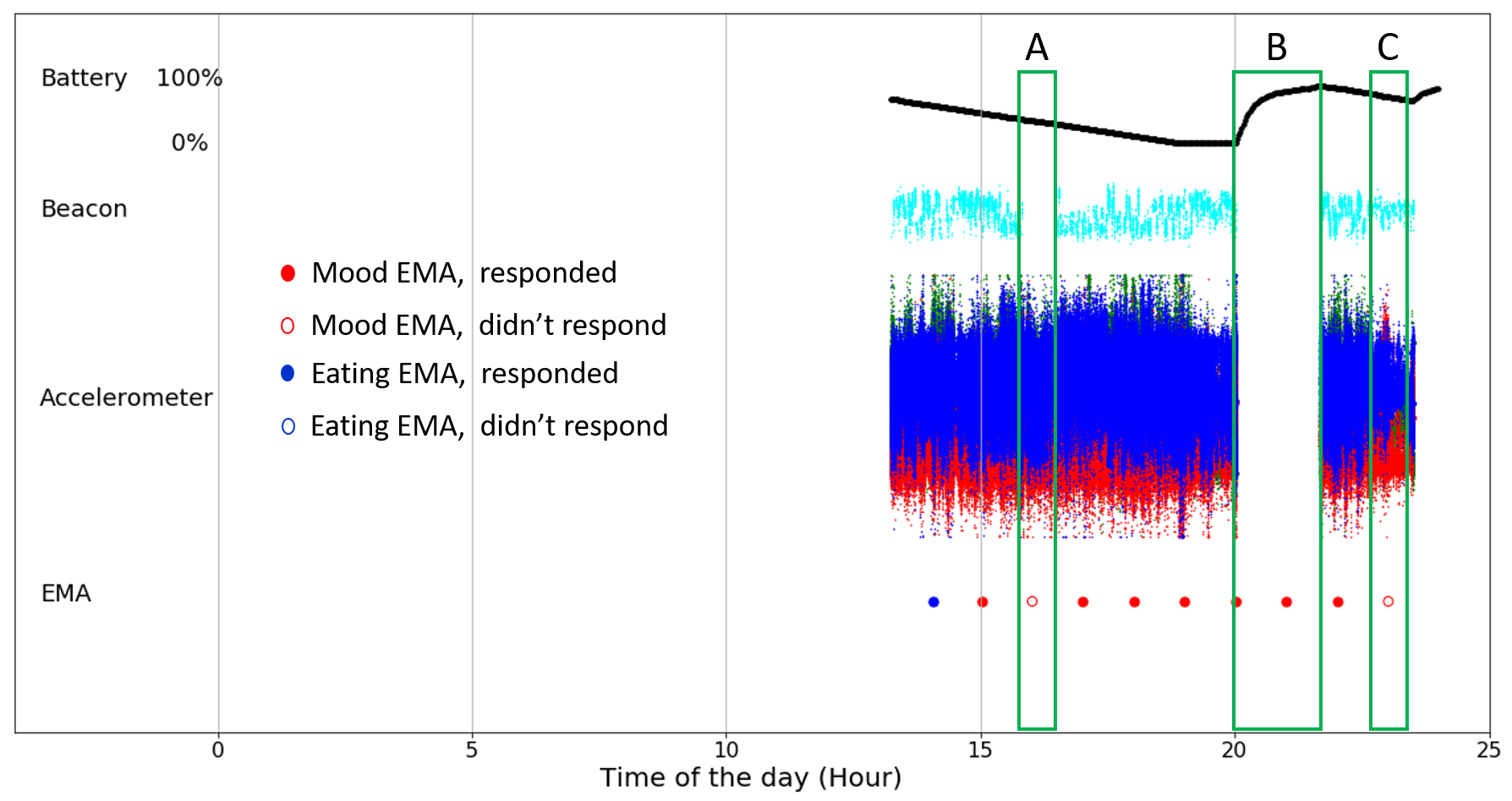}
  \caption{$Scenario\ A$: accelerometer data is available, but there is no beacon data. The participant did not answer to the EMA sent during that time, and so it cannot be determined whether the participant was at home or not. $Scenario\ B$: the watch was being charged, and there was no accelerometer and beacon data. However, the participant answered to the EMAs. So, the participant was at home during that time. $Scenario\ C$: the participant is at home because both beacon and accelerometer data are available, but he/she did not answer to the EMA sent during that time.}
  \label{fig:daylong_sub1}
\end{subfigure}

\begin{subfigure}[b]{.9\textwidth}
  \centering
  \includegraphics[width=.9\linewidth]{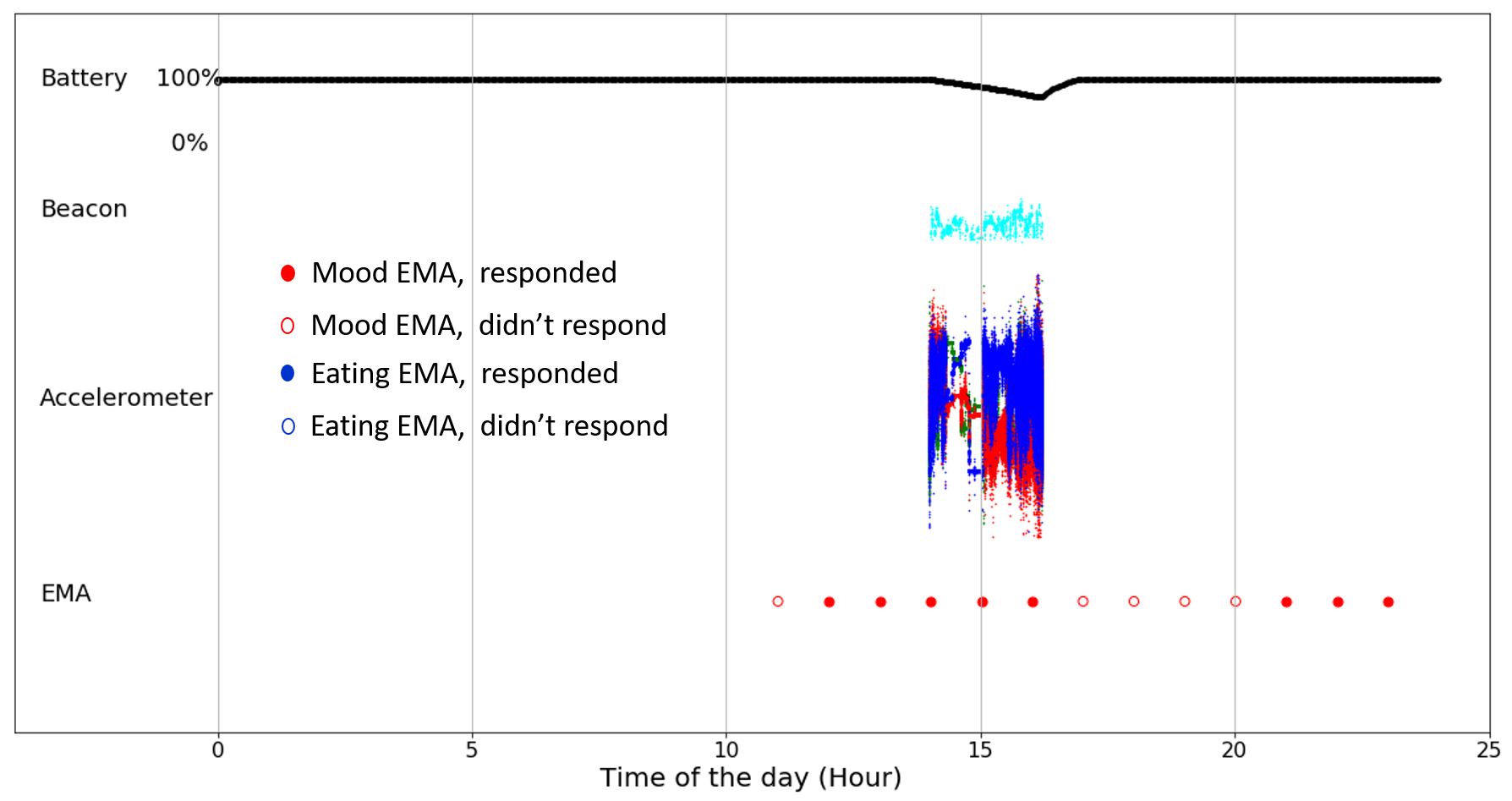}
  \caption{A scenario where the participant wore the watch for about 2.5 hours, and then began charging it. The participant answered all the EMAs while wearing the watch, but then did not answer several other EMAs.}
  \label{fig:daylong_sub2}
\end{subfigure}
\caption{Daylong data of battery percentage, beacon RSSI, accelerometer and EMA responses for two days from two different participants.}
\label{fig:daylong_data}
\end{figure}

Overall, the system sent a total of 14413 EMAs to the participants with 13776 and 637 EMAs for mood and eating, respectively. The participants responded to 4750 of the EMAs (4224 for mood and 526 for eating). Currently, MFED does not localize the participants or the study phones in real time. The EMAs are sent regardless of the location of the participants or the study phones. So, there are instances where a user might be out of home or away from his/her study phone when an EMA is sent. To better explain the EMA response rate and as a clarifying example, the daylong data for two days from two different participants are shown in Figure \ref{fig:daylong_data}. We see that the participants wore the watch for a part of the days, which is typical as they spend time outside of home for many purposes including work, school, sports, and shopping. Also, the watches might run out of energy while they are being used. In Figure \ref{fig:daylong_data}(a), we see that the participant started wearing the watch after the noon, and there is a gap when the watch was being charged. The red and blue circles represent mood and eating EMAs, respectively, and the filled circles represent the EMAs that the participants answered. We see that the participant (Figure \ref{fig:daylong_data}(a))  responded to all the EMAs that day except two. Three scenarios are marked in the figure. For scenario A, accelerometer data is available, but there is no beacon data. We see that the participant did not answer to the EMA sent during that time. We do not know whether the participant was at home or not because the participant was either outside of home or somewhere in the home where the beacon signals were not available. During scenario B, the watch was being charged, and there was no accelerometer and beacon data. However, the participant answered to the EMAs. So, the participant was at home during that time. For scenario C, we know that the participant is at home because both beacon and accelerometer data are available, but he/she did not answer to the EMA sent during that time. In Figure \ref{fig:daylong_data}(b), the participant wore the watch for about 2.5 hours, and then began charging it. The participant answered all the EMAs while wearing the watch, but then did not answer several other EMAs. In this case, we don't know whether he/she was at home when the EMAs are not answered. At night, the participant started answering the EMAs, and it indicates he/she was at home during that time but forgot to wear the watch. 

The participants replied to about 31\% of the hourly EMAs. Though the response rate is low, it is reasonable and expected. People spend a significant amount of time outside their home. The participants of our study do not carry the study phone outside home, and  thus cannot respond to the EMAs that time. Our system does not track whether a participant is at home or not. Using the smartwatches for such tracking is not practical as users do not wear watches all the time for different reasons including for charging and forgetting to wear the watch, as illustrated above. Also, users do not usually wear the watch when they are out of home. Since we do not know the location of the user, we send the mood EMAs every hour regardless of the users location. So, the response rate to the hourly EMAs is expected to be low. 

\begin{figure}\centering
\includegraphics[width=1\linewidth]{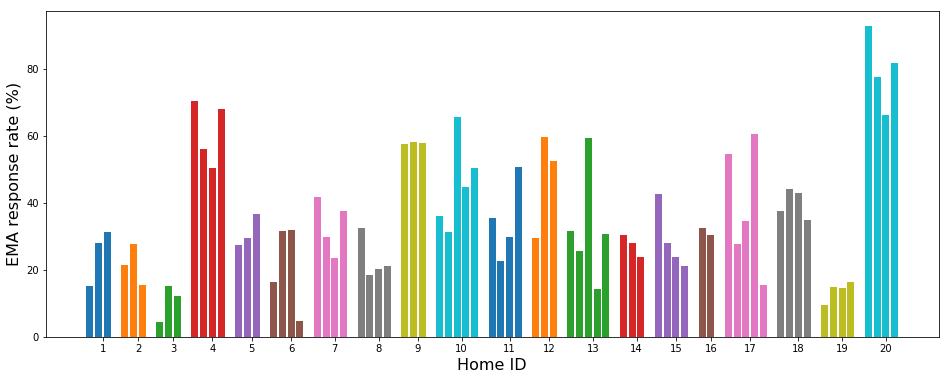}
\caption{EMA response rate for each of the participants that are grouped with same color for the same family.}
\label{fig:ema_rate_person}
\end{figure}

\begin{figure}
\centering
\begin{subfigure}{.5\textwidth}
  \centering
  \includegraphics[width=.9\linewidth]{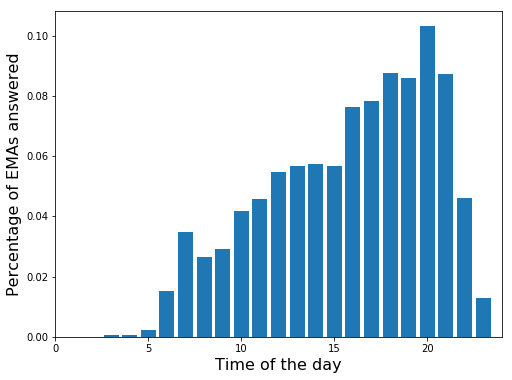}
  \caption{}
  \label{fig:tod}
\end{subfigure}%
\begin{subfigure}{.5\textwidth}
  \centering
  \includegraphics[width=.9\linewidth]{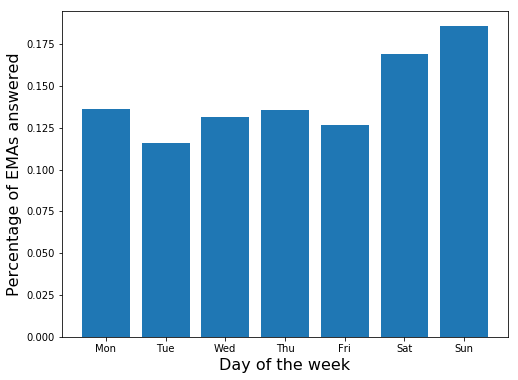}
  \caption{}
  \label{fig:dow}
\end{subfigure}
\caption{EMA response rates at different (a) time of the day and (b) days of the week.}
\label{fig:tod_dow}
\end{figure}

The response rate for the EMAs differs from person to person. Figure \ref{fig:ema_rate_person} shows the response rates for each of the participants that are grouped with same color for the same family. We see that the response rates differ significantly from family to family. However, the correlation coefficient between the response rates of the participants and the mean response rate of the corresponding homes is 0.87. This indicates that the response rates among family members are highly correlated. The rates of EMA responses differs at different times of the day as depicted in Figure \ref{fig:tod}. There are more responses in the evening compared to morning and noon. This might be because people are less likely to be at home during the morning and noon than evening for reasons including work and school.  Figure \ref{fig:dow} shows the EMA response rates for different days of the week. It shows that participants responded to EMAs more during the weekend than the weekdays. This may be because people are more likely to be at home or less likely to be busy during the weekends.

\subsubsection{\textbf{Eating Events:\\}}
The participants answered to 526 of the eating EMAs out of the 637 sent.  They responded that they were eating for 383 of the EMAs (256 meals, 87 snacks, 28 drinks, and 12 undefined). In cases where the participants were not eating, they were asked about what they were doing. There were five options available to the users: 1) Using a personal phone, 2) Smoking, 3) Fixing hair, 4) Putting Sunscreen or lotion, and 5) Other (Open text area). The participants could select one or more of the options and could  provide text inputs in the open text area. The participants selected only one option for all the EMAs except for two EMAs where they selected 2 ($Using\ my\ phone$ and $Other$ ) and all the 5 options. The response count for different activities are listed in Table \ref{table:ema_not_eat}. It shows that using the phone is one of the most confounding gestures for eating. The most frequent activities mentioned in the open text area include using a computer/laptop and Watching TV/movies. There are wide variety of activities that were mentioned in the open text area. It indicates that many activities found in the wild confounds with eating. The responses provided by the users can be used as ground truth for activity recognition tasks, particularly for eating activity detection. 

\begin{table}
\begin{tabular}{|c|c|}
\hline
\textbf{Activities} & \textbf{Count} \\ \hline
Using personal phone & 67\\ \hline
Smoking & 2 \\ \hline
Fixing own hair & 4 \\ \hline
Putting on sunscreen or lotion & 4 \\ \hline
Other & 71 \\ \hline
\end{tabular}
\caption{Activities during falsely detected eating events}
\label{table:ema_not_eat}
\end{table}

\begin{figure}\centering
\includegraphics[width=0.6\linewidth]{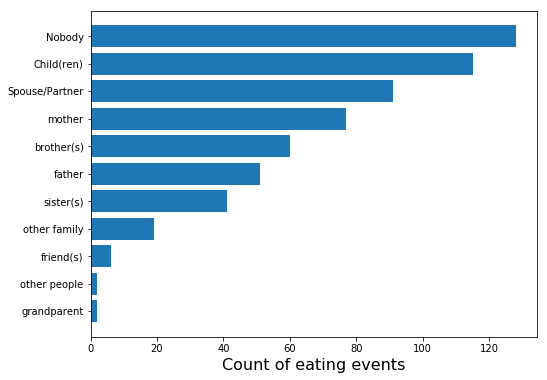}
\caption{Frequency of other family members or other persons eating with the participant who responded to the eating EMA.}
\label{fig:with_whom_count}
\end{figure}

\begin{figure}\centering
\includegraphics[width=0.6\linewidth]{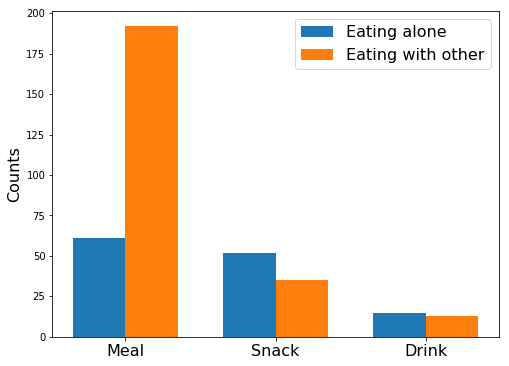}
\caption{Counts of meals, snacks and drinks while eating alone and eating with others.}
\label{fig:with_whom_breakdown}
\end{figure}

The participants were asked about other persons who were eating with them. The options available for this query are Nobody, Spouse/Partner, Child(ren), Mother, Father, Sister(s), Brother(s), Grandparent, Other family, Friend(s) and Other people. Figure \ref{fig:with_whom_count} lists the frequency of the family members or others present during eating. It shows that both eating alone and eating with family members are common at homes. Figure \ref{fig:with_whom_breakdown} shows the breakdown of meals, snacks and drinks during eating alone and eating with others. It depicts a phenomenon of eating at home - people eat meals together more than alone and eat snacks alone more than with others.          

\subsubsection{\textbf{Collaborative Ground Truth:\\}}
We use the family role of each of the participants and the relationship between the participants for collaborative ground truth. For example, when a child in a family answers that she is eating with her mother, it provides ground truth for the mother. If the system fails to detect that eating event for the mother or she does not respond to her eating EMA, the response from the child can be used as ground truth for the mother.

There are 240 EMA responses where the participants mentioned that they were eating with others. As mentioned earlier, all members were not interested in participating in the study for some families. There are 68 EMA responses where the participants mentioned they ate with some family members (spouse/partner, child, mother, father, brother, sister) who didn't participate in the study. There are 101 EMA responses where the participants mentioned the relationship with other participants she was eating with, but the other participant cannot be unambiguously identified because multiple similar participants are available at that home. For example, when a mother said that she is eating with her children, and there are more than one child in the home, we cannot unambiguously determine which child she was eating with. In 165 of the EMA responses the participants mentioned about a total of 257 other participants who can be detected unambiguously. However, multiple participants may mention about one participant. For example, two children may mention that they were eating with their mother. In such case, we get the ground truth for the mother from two sources. There are 35 such instances, and so we get ground truth for 222 instances through the collaborative  approach. We considered a time window of 15 minutes for this purpose. 56 instances also have ground truth through the first-person approach where the other participants responded to their EMA survey. So, the collaborative approach gives ground truth for 170 instances for which first-person ground truth is not available. There are several reasons for the absence of first-person ground truth for eating events. Firstly, the participant might not wearing the watch during the meal, and so no eating event was detected for that participant. Such scenarios are common when the watch is being charged or the participants forget to wear the watch. Secondly, the system might have detected the eating event, but the participant did not respond to the EMA. It might happen when the phone is not near the participant and he/she does not see the notification. Thirdly, our system might fail to detect that eating event for the participant.








\section{Related Works}\label{sec:related_works}
Obesity, diabetes, cardiovascular diseases, and many other human health applications motivate the research in the area of automated dietary monitoring and eating detection. Consequently, most of the related works in this area attempt to estimate energy or calorie intake, characterize food, and quantify mass intake by monitoring the eating gestures or sips as well as the eating events such as meal or snack~\cite{fontana2015detection,thomaz2017challenges}. State-of-the-art approaches have been using various wearable sensing modalities to detect either eating gestures or eating events. These approaches vary not only on the employed sensing modalities but also on the placement of such sensors on the body as well as the signal processing, classification, and learning approaches~\cite{vu2017wearable,hassannejad2017automatic,schiboni2017automatic}.

Advances in imaging technologies have provided the opportunity to continuously carry a portable or wearable camera which can capture dietary moments. With image processing techniques such as normalized cut based segmentation and SVM based classification~\cite{zhu2010image}, scale invariant feature transform based 3D reconstruction~\cite{kong2012dietcam}, or cloud-based human computing platforms~\cite{thomaz2013feasibility,thomaz2013technological}, images captured from a hand-held, or on-body camera have been used to assess users dietary behavior. These approaches show the potential in addressing multiple aspects of dietary applications, such as ingested mass quantification, food intake characterization, as well as energy intake estimation. Yet, privacy concerns have restricted the ability of these approaches to be extended to real-world deployments~\cite{thomaz2017challenges,hassannejad2017automatic}.

Acoustic signals of chewing and swallowing have also been investigated as indicators of dietary activity. In early attempts, miniature microphones were placed inside the ear canal as earplugs, which acquired acoustic signal related to eating episodes, and those signals were filtered and classified using CART, NB, KNN, and HMM classifiers to detect mastication and swallowing in controlled settings~\cite{amft2005analysis,amft2006methods}. These approaches showed high recognition rate (upto 99\%), but only on a handful of foods (e.g. potato chips, apple) with a small population (4-5 subjects) in controlled settings. A throat microphone was used to capture better swallowing sound in addition to a microphone in the ear canal for chewing sounds to collect a total of 64.5 hours of data~\cite{sazonov2008non,sazonov2010automatic,lopez2012automatic}. These data were used for eating episode detection by using frequency and wavelet features, ans well as clustering and affinity propagation was used to estimate counts of food items in a certain episode. In real-world settings, the acoustic signal becomes corrupted by environmental noise and surrounding sounds, and thus hampers the performance of those audio-based methods. To avoid the impacts of environmental noise, methods have been proposed attempting various noise subtraction algorithms based on energy and spectral properties of the eating related sounds and the environmental noise collected using two independent microphones packaged as a hearing aid~\cite{passler2011acoustical,passler2012food,passler2014food}. 
With similar motivation, a uni-directional microphone based neckpiece was designed to capture acoustic signals from the throat while reducing external noise, and was able to recognize twelve activities at 79.5\% F-score in lab setting and four activities at 71.5\% F-score in realistic setting~\cite{yatani2012bodyscope}. With assumptions that a body-worn piezoelectric microphone would be less affected by environmental noise, a neckpiece sensor was designed using such microphone, yet didn't take off for real-world evaluation of eating episode detection~\cite{rahman2014bodybeat,rahman2016predicting}. 
A wrist-worn microphone was used in a semi-controlled study to acquire data around eating episodes and used random forest classification on clustered audio frames to achieve 79.8\% F-score~\cite{thomaz2015inferring}. 
Other modalities such as electrogluttography~(EGG) and electromyography~(EMG) have also been proposed to capture jaw motion and swallowing events in a similar manner~\cite{sazonov2012sensor,farooq2014novel,zhang2016diet}.
Also, combining audio with images from an ear-worn device were proposed for both eating gesture detection and food identification based energy intake estimation~\cite{liu2012intelligent}.
These above mentioned works demonstrate that audio based systems can acquire good performance in detecting eating gestures such as eating gesture and ingestion in controlled environments, but fail to translate those performance in real-world scenarios~\cite{kalantarian2016comparison,merck2016multimodality}.

Human motion sensing has shown notable potential in inferring various activities of daily living including eating~\cite{junker2008gesture}. Specially with the recent proliferation of such sensing modalities in pervasive and ubiquitous computing devices, such sensors require much less effort to deploy in real-world while providing signals with reliable quality. Consequently, many researchers have attempted to use motion as a supplementary modality for detecting eating activity.
Motion signal from lower and upper arm, audio from ear canal and neck collar, and EMG from throat were combined to perform feature distance based classification of eating events~\cite{amft2008recognition}. Wrist motion was used in combination with chewing and swallowing sound to acquire better performance in eating gesture detection and food characterization~\cite{fontana2014automatic,fontana2015detection}. While applying pattern recognition with these sensor streams provide around 80\% accuracy in controlled settings, such methods were not evaluated in long-term free-living settings. \cite{ye2015automatic}~proposed a dual location motion based system by using both smart watch and Google Glass accelerometers, but evaluated on a controlled protocol of activities. Similarly, motion data from head movement was captured using a glass-mounted accelerometer and KNN classification was performed on small amount of lab collected data~\cite{farooq2018accelerometer}.

Considering the importance of data collection from the real-world, recent approaches are focusing more on wrist-worn motion sensing due to its unobtrusiveness and higher user conformance. Focusing on energy intake estimation, a rule-based eating gesture counting algorithm and a wrist worn sensing device named eating gesture counter were proposed~\cite{ramos2015improving,shen2017assessing,muth2017designing}. It was evaluated in two controlled settings - a lab setup and an instrumented cafeteria; performance evaluations in general scenarios require further attention. Also, the participants had to manually press a button to start the data collection and eating gesture counting. Among other approaches for eating gesture detection, motif based template matching based pattern segmentation followed by Random Forest classification were evaluated in semi-controlled environments~\cite{zhang2018sense}. Similarly, a micro-movement segmentation based HMM classifier was designed to detect eating gestures in a known eating episode~\cite{kyritsis2017automated}. These approaches highlight the need for an automated meal detection method for those to be applicable in real-world scenarios.

With the motivation for free-living data collection and application, Dong et al. attempted to develop an eating period detection algorithm \cite{dong2012new, dong2014detecting}. They strapped a smart phone on the wrist of the user's dominant hand to utilize the accelerometer and the gyroscope of the phone and collected data for one-day per participant from 43 participants in free-living settings. The proposed algorithm was based on an assumption that eating periods are preceded and succeeded by vigorous motion patterns, which is too simplistic against real-world confounding gestures. Though this work found correlation between energy intake and eating gesture counts, but was not evaluated against long-term inter- and intra-person variations and uncertainty.

Recent approaches are emerging to translate the in-lab high performance systems to out-of-lab realistic long-term setting. For example, lab data collected from 20 subjects wrist-worn motion sensor were used to train models which were evaluated in free-living setting on 7 participants each for 24 hour and on 1 participant for a month~\cite{thomaz2015practical}. To acquire ground truth on food intake, a first-person point-of-view camera was worn by the participants. Even though the data lacks inter- and intra-person variation in eating behavior; yet this was a major step toward implementing realistic solution that can be used by various dietary monitoring applications. Following that trend, Mirtchouk et al.~\cite{merck2016multimodality,mirtchouk2017recognizing} have conducted an investigative study on eating recognition using head- and wrist-worn motion sensor and an ear-canal audio sensor for 12 participants in-lab and out-of-lab settings. Their contributed ACE~(accelerometer and audio-based calorie estimation) dataset contains 6 participants total 12 hour data in lab setting, ACE-free-living dataset contains data for 5 of those ACE participants one-day each data in real-world setting, and ACE-external dataset contains data for 6 new participants, 5 of which for 2 days and the 6th one for 5 days in free-living setting~\cite{acedataset}. Keum et al. \cite{chun2018detecting} instrumented a necklace with proximity sensors that detect eating through sensing head and jawbone movements. They evaluated the performance of the system both in controlled setting, and in free-living context. Though performance drops significantly in the free-living context compared to the lab study, the results show the promise of using such devices for real-world deployments. Earbit \cite{bedri2017earbit} is a head-mounted wearable system for detecting eating episodes. Though EarBit can collect data from multiple sensors (inertial, acoustic, and optical), the study shows that two inertial sensors, one behind the ear and the other behind the neck are more effective than other sensing modalities. Results from the study show that EarBit is effective in detecting chewing and eating episodes in an unconstrained environment. 

Ecological Momentary Assessment (EMA) is widely used to understand eating behavior and the influence of different factors on eating. Julia et al. \cite{reichenberger2018no} used EMA to find the effect of stress, negative and positive emotions on eating behavior, particularly on taste- and hunger-based eating. The study shows that stress and emotions influence eating behavior significantly. For example, higher stress reduces but positive emotion increases taste-eating. Genevieve et al. \cite{dunton2017daily} used EMA to find association between stress and eating in mother-child dyads. EMA was used to collect information on perceived stress, and healthy/unhealthy food consumption. Results show that healthy and unhealthy eating by the children are coupled with those by the mothers' at the day level. It depicts the effect of family members on each other in terms of eating habits. Andrea et al. \cite{goldschmidt2017contextual} used EMA to collect data from adults with obesity and studied the association between contextual factors and eating in the absence of hunger (EAH). The study reports that there is lack of hunger in 21\% of eating events, and the participants perceived overeating for these events. Khouloud et al. \cite{alabduljader2018ecological} developed a smartphone EMA app to record data related to wanting and liking of food. Results show less food wanting and lower intensity of food liking among the adults with more body fat.

\section{Discussion and Future Work}\label{sec:discussion}
MFED is designed and developed to capture a wide range of information related to family eating dynamics. The broader purpose is to build dynamic and networked models of FED, and use these models to drive personalized, adaptive and just-in-time interventions that have potential to be effective in the long term modification of eating behavior and prevention of obesity. Modeling FED, designing interventions or validating the effectiveness of FED approach are beyond the scope of this paper. The goal of this work is to build a foundation that would support future endeavors related to family eating dynamics. EMA is not associated with the localization of the participants, and this work does not address the problem of localization. In future, we will incorporate localization features in MFED. The beacon RSSI data collected from the deployments will be helpful to develop solutions for in-home localization of the users.

In contrast to lab-studies, real-world deployments need to address many challenges including usability, user convenience, and resource constraints of the devices. A wristwatch is a very common personal device, and there is almost no inconvenience in using it. Considering all the issues, MFED uses a single smartwatch for eating detection, though detecting eating activities in the real-world using a smartwatch has proven to be challenging \cite{thomaz2015practical, mirtchouk2017recognizing, zhang2017generalized}. Additionally, MFED requires detecting the eating events in real-time. We designed the system to address these challenges and incorporated features to reduce burden on the users. For example, the watch app starts automatically when the watch restarts, and so the users do not need to start the app. In fact, the users do not interact with the watch app at all. It runs seamlessly without user intervention.   

Before real deployments, we collected data from lab settings to build eating gesture classification models. We also included data for non-eating activities from free-living context. The eating gestures associated sensor data differ significantly based on several factors including the type of food, utensils (e.g., spoon, fork, chopstick, bare hand) used to eat the food, body postures, context, and individual differences. These factors along with confounding gestures from non-eating activities make eating gesture detection challenging, particularly in the the wild. Though the lab data captures a wide range of gestures, they are still limited compared to the diversity of gestures in the real-world. The data we collected from the deployments provide ground truth for many eating events, most importantly from real home contexts. Though there is no ground truth available for individual eating gestures, the ground truth for eating events can be exploited with semi-supervised methods to build more robust and accurate eating gesture detection models.   

In addition to eating gesture detection, it is also challenging to define an eating event from detected eating gestures. The intervals between eating gestures are usually irregular and depend on factors like context, food type, and habits. In addition to irregular intervals, the accuracy limitation of the eating gesture classification model makes it more challenging to detect the eating events. We use a heuristic as proposed by Mirtchouk et al. \cite{mirtchouk2017recognizing} that clusters the eating gestures within one-minute intervals, and then we detect eating events using these clusters. Future work includes using the data from the deployments to better understand the structures of eating events, and to develop better methods for eating event detection. Personalized models usually work better than general models, and in the future, we will incorporate personalized models in MFED for eating gesture and eating event detection. 

We did not attempt to keep journals or ask participants to fill in missed meals. So, it is not possible to detect the false negatives as our system fails to detect the eating events. We get only the true positives and the false positives. Any model that increases true positives and decreases false positives for this data is likely to decrease false negatives. Future works include using the data to develop more accurate models that reduces false positives. Though it is not possible to verify the false negatives, if a model detects eating events with very few false positives, the newly detected eating events (that were not detected during deployment) can be used to estimate the proportion of false negatives. The data we collected will facilitate such research in the future.


The optimal frequency with which to assess stress and mood via ecological momentary assessment, with regard to both accuracy and compliance, is still an open question in the field of behavioral science. The data we collected in this study will facilitate future research in this area. However, to balance the trade-off between collecting more temporally granular data and the convenience of the users, we chose 1-hour intervals for measuring within-subject changes in mood and stress. Future works include finding out better timing and frequency for EMA surveys using the data collected from this study. Technologies for automatic detection of mood and stress, particularly using wearables \cite{sandulescu2015stress, zenonos2016healthyoffice} are advancing. In order to reduce the burden on the users and to gather more temporally granular data on mood and stress, future works should focus on incorporating automatic solutions for mood and stress detection in MFED, ensuring the privacy of the users. Due to the small form factor, smartwatches are not suitable for EMA surveys. However, voice-based interactive systems like MedRem \cite{mondol2016medrem} can be developed in the watch for EMA purposes. In the future, we will explore the feasibility of using such a solution for MFED.  

Though we run the model in the base station, energy is still a critical issue for the watch because we detect eating in real time, and streaming data continuously to the base station would drain significant energy from the watch.  Computation and memory are not critical for eating detection on the base station, but they are critical for the watch. However, the base-station detects eating from all the watches of the corresponding home in real-time. So, an efficient method allows us to use low-cost device as a base-station. Since our method requires low computation and memory, it can be used in further works that would attempt on watch eating detection.

The collaborative ground truth is more feasible for home than outside because there are usually few people in a home, and the relationship among participants are defined. For example, when a daughter confirms that she is eating with his father, we know who her father is. Since a person can eat outside of home with many different people with different relationships, collecting such ground truth would be relatively more complex. However, the collaborative ground truth has potential applications for other home based applications beyond MFED and also for other settings where the relationship between the participants can be defined.

The dataset collected from the deployments consists of accelerometer, battery and beacon readings from smartwatches as well as EMA responses from the users. This is a unique dataset that will be made public, and it would be invaluable for future research related to family eating dynamics and other family based systems. The dataset can be used in building better FED monitoring systems as well as new, dynamic and networked models of FED that will be able to drive real-time, in the wild interventions. Approaches based on family eating dynamics have the potential to be very effective in addressing obesity. This paper lays the foundation that would support future endeavors to tackle the obesity problem using FED.

\balance
\bibliographystyle{ACM-Reference-Format}
\bibliography{references}

\end{document}